\documentclass[twocolumn]{aastex63}

\usepackage{xspace}
\usepackage{tablefootnote}

\newcommand{\OII}{[\ion{O}{2}]\xspace}
\shorttitle{Dark Energy Explorers}
\shortauthors{Lindsay R. House}

\begin{document}

\title{Using Dark Energy Explorers and Machine Learning to Enhance the \\ Hobby-Eberly Telescope Dark Energy Experiment}

\author[0000-0002-1496-6514]{Lindsay R. House}
\altaffiliation{NSF Graduate Research Fellow}
\affiliation{Department of Astronomy, The University of Texas at Austin, 2515 Speedway Boulevard, Austin, TX 78712, USA}
\correspondingauthor{Lindsay R. House}
\email{lindsay.r.house@gmail.com}

\author[0000-0002-8433-8185]{Karl Gebhardt}
\affiliation{Department of Astronomy, The University of Texas at Austin, 2515 Speedway Boulevard, Austin, TX 78712, USA}

\author[0000-0003-0792-5877]{Keely Finkelstein}
\affiliation{Department of Astronomy, The University of Texas at Austin, 2515 Speedway Boulevard, Austin, TX 78712, USA}

\author[0000-0002-2307-0146]{Erin Mentuch Cooper}
\affiliation{Department of Astronomy, The University of Texas at Austin, 2515 Speedway Boulevard, Austin, TX 78712, USA}
\affiliation{McDonald Observatory, The University of Texas at Austin, 2515 Speedway Boulevard, Austin, TX 78712, USA}

\author[0000-0002-8925-9769]{Dustin Davis}
\affiliation{Department of Astronomy, The University of Texas at Austin, 2515 Speedway Boulevard, Austin, TX 78712, USA}

\author[0000-0002-1328-0211]{Robin Ciardullo}
\affiliation{Department of Astronomy \& Astrophysics, The Pennsylvania State University, University Park, PA 16802, USA}
\affiliation{Institute for Gravitation and the Cosmos, The Pennsylvania State University, University Park, PA 16802, USA}

\author[0000-0003-2575-0652]{Daniel J Farrow}
\affiliation{Centre of Excellence for Data Science, Artificial Intelligence and Modelling (DAIM), University of Hull, Cottingham Road, Kingston-upon-Hull HU6 7RX, UK}

\author[0000-0001-8519-1130]{Steven L. Finkelstein}
\affiliation{Department of Astronomy, The University of Texas at Austin, 2515 Speedway Boulevard, Austin, TX 78712, USA}

\author[0000-0001-6842-2371]{Caryl Gronwall}
\affiliation{Department of Astronomy \& Astrophysics, The Pennsylvania State University, University Park, PA 16802, USA}
\affiliation{Institute for Gravitation and the Cosmos, The Pennsylvania State University, University Park, PA 16802, USA}

\author[0000-0002-8434-979X]{Donghui Jeong}
 \affiliation{Department of Astronomy \& Astrophysics, The Pennsylvania State University, University Park, PA 16802, USA}
 \affiliation{Institute for Gravitation and the Cosmos, The Pennsylvania State University, University Park, PA 16802, USA}

\author[0000-0001-6421-0953]{L. Clifton Johnson}
\affiliation{Adler Planetarium, 1300 S. DuSable Lake Shore Dr., Chicago, IL 60605, USA}
\affiliation{Center for Interdisciplinary Exploration and Research in Astrophysics (CIERA) and Department of Physics and Astronomy, Northwestern University, 1800 Sherman Ave., Evanston, IL 60201, USA}

\author[0000-0001-5561-2010]{Chenxu Liu}
\affiliation{South-Western Institute for Astronomy Research, Yunnan University, Kunming, Yunnan, 650500, People’s Republic of China}
\affiliation{Department of Astronomy, The University of Texas at Austin, 2515 Speedway Boulevard, Austin, TX 78712, USA} 

\author[0000-0002-0977-1974]{Benjamin P. Thomas}
\affiliation{Department of Astronomy, The University of Texas at Austin, 2515 Speedway Boulevard, Austin, TX 78712, USA}

\author[0000-0003-2307-0629]{Gregory Zeimann}
\affil{Hobby Eberly Telescope, The University of Texas at Austin, Austin, TX, 78712, USA}

\begin{abstract}

We present analysis using a citizen science campaign to improve the cosmological measures from the Hobby-Eberly Telescope Dark Energy Experiment (HETDEX\null). The goal of HETDEX is to measure the Hubble expansion rate, $H(z)$, and angular diameter distance, $D_A(z)$, at $z = 2.4$, each to percent-level accuracy. This accuracy is determined primarily from the total number of detected Lyman-$\alpha$ emitters (LAEs), the false positive rate due to noise, and the contamination due to \OII emitting galaxies. This paper presents the citizen science project, \href{https://www.zooniverse.org/projects/erinmc/dark-energy-explorers}{\textit{Dark Energy Explorers}}, with the goal of increasing the number of LAEs, decreasing the number of false positives due to noise and the \OII galaxies. Initial analysis shows that citizen science is an efficient and effective tool for classification most accurately done by the human eye, especially in combination with unsupervised machine learning. Three aspects from the citizen science campaign that have the most impact are 1) identifying individual problems with detections, 2) providing a clean sample with 100\% visual identification above a signal-to-noise cut, and 3) providing labels for machine learning efforts. Since the end of 2022, \textit{Dark Energy Explorers} has collected over three and a half million classifications by 11,000 volunteers in over 85 different countries around the world.  By incorporating the results of the \textit{Dark Energy Explorers} we expect to improve the accuracy on the $D_A(z)$ and $H(z)$ parameters at $z = 2.4$" by $10-30\%$. While the primary goal is to improve on HETDEX, \textit{Dark Energy Explorers} has already proven to be a uniquely powerful tool for science advancement and increasing accessibility to science worldwide.

\end{abstract}


\section{Introduction}
\label{sec:intro}

Supernovae observations discovered that the universe is undergoing an accelerated expansion (\citealt{Riess98, Perlmut, Riess21}), which has been confirmed by a myriad of follow-up cosmological observations (\citealt{2dFSurvey, SDSSmain, WMAP, plank, DES}). The community is struggling for a theoretical understanding of this acceleration \citep{kolb}, with the cosmological constant as the primary culprit \citep{Weinberg}. There are experiments \citep{DES, DESI, Euclid} to measure this acceleration both with an improved accuracy and longer-time baseline. Both improvements will help limit the available physical models for explaining the accelerated expansion (i.e. dark energy).

The initial experiments measure the accelerated expansion during the late-time of the universe at $z < 1$. While planned experiments like DESI and Euclid provide significant increase in the accuracy of the expansion rate, they will have only a modest increase in redshift range \citep{DESI, Euclid}. Currently, the uncertainties on $H(z)$ and $D_A(z)$ at different redshifts range from 1.8\%–3\% (see the summary in \cite{DESI}). The Baryon Oscillation Spectroscopic Survey (BOSS) and its extension, eBOSS, measure $D_{\mathrm{A}}$ with accuracy of 1.8\% at $z = 0.5$ \citep{bautista, gil-marin}, 2.0\% at $z = 0.7$ \citep{deMattia}, and $\sim$ 3\% at $z = 2.3$ \citep{deMasDesBourboux}, and the Dark Energy Survey (DES) gives an uncertainty of 2.7\% at $z = 0.84$ \citep{DES}. Various missions, such as the Dark Energy Spectroscopic Instrument (DESI) and Euclid are expected to achieve a precision of $\sim$ 0.5\% at $z \sim$ 1 (\citealt{DESI, Euclid}).

In order to detect any evolution in the nature of dark energy over cosmic time, it is necessary to cover as large a redshift range as possible. The Hobby-Eberly Telescope Dark Energy Experiment (HETDEX) is designed to study the expansion rate at $1.9 < z < 3.5$ with an accuracy comparable to even the best low-$z$ experiments \citep{Gebhardt21}. HETDEX will determine redshifts of at least one million Lyman-$\alpha$ emitting (LAE) galaxies from $1.9 < z < 3.5$. Our approach is to observe an area of 540 square degrees with an instrument composed of 74 integral field units feeding 156 spectrographs, over a spectral range of 350-550 nm. 

HETDEX will use these million LAE galaxies to measure and analyze the full-shape of the galaxy power spectrum
and galaxy correlation function, including Baryon Acoustic Oscillations (BAO), of the $1.9 < z < 3.5$ universe to determine the epoch's dark energy density. Measurements of clustering in the directions parallel and perpendicular to the line of sight provide constraints on the Hubble parameter, $H(z)$, and the angular diameter distance, $D_{\mathrm{A}}(z)$ through the Alcock–Paczynski test \citep{AlcockPacz, Sanchez14}. The angle-averaged clustering measurements determine the average distance, $D_{\rm V}(z)$, as

\begin{equation}
    D_{\mathrm{V}}(z) \propto \frac{D_{\mathrm{A}}(z)^2}{H(z)}.
\end{equation}
With the expected volume and density of the measured LAEs, HETDEX will determine the Hubble expansion rate, $H(z)$, and the angular diameter distance, $D_{\mathrm{A}}(z)$, each to $0.8\%$ around $z = 2.4$, which would be the most accurate measure of the Hubble expansion at this epoch \citep{Gebhardt21}. These accuracies translate to an overall accuracy on the volume averaged distance, $D_{\mathrm{V}}$, below 0.7\%.

\begin{figure*}[ptb]
    \begin{center}
         \includegraphics[scale=0.5]{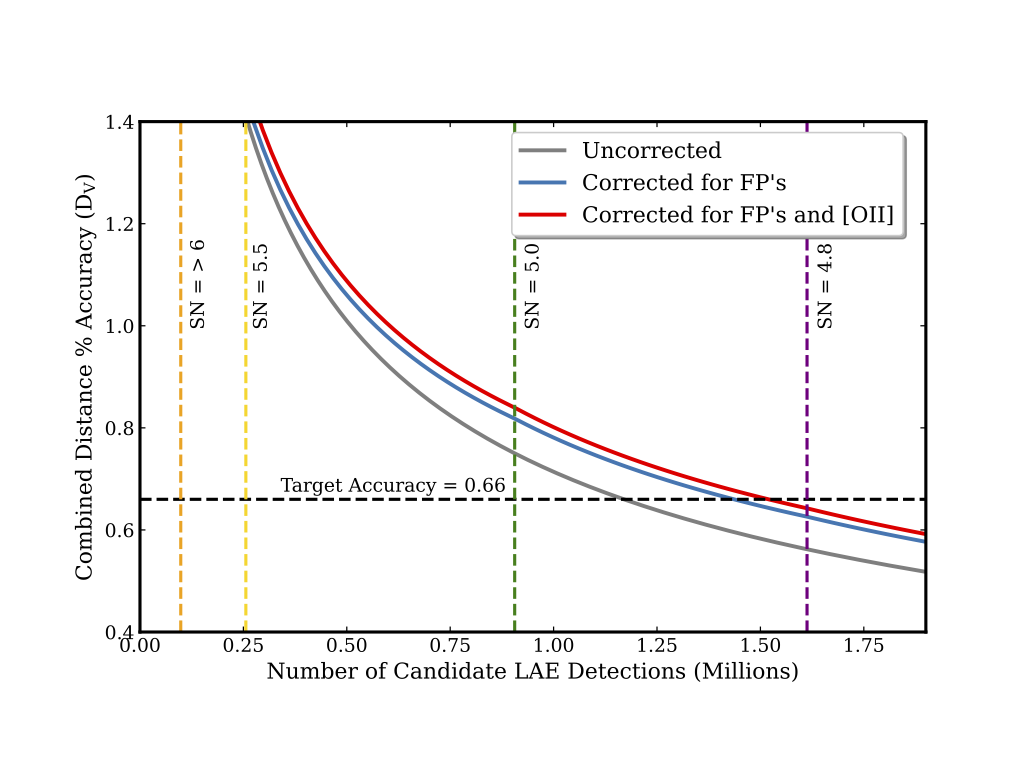}
  \caption{Based on the modelling of HDR2 results, this plot shows the accuracy of $D_{\mathrm{V}}$ as a function of the number of detected LAEs and signal-to-noise. The primary relation is that $D_{\mathrm{V}}$ goes as the inverse square root of the number of LAEs, and this is given by the gray line assuming no corrections for false positive rate or \OII contamination. This demonstrates how the accuracy of the cosmological parameter, $D_{\mathrm{V}}$ is affected by false positives (FP), as the blue line, and further including the \OII contamination as the red line. The black horizontal line represents the target goal for the whole HETDEX survey. The precision achieved for a given signal-to-noise limit is given by the intersection of the gray, blue, and red curves with the vertical dashed lines that represent the different signal-to-noise cuts. Note that the blue and red curves are less accurate due to contamination. Here, we assume a galaxy bias of $b_{\rm LAE} = 2$. The S/N cut and/or the number of LAEs that is required to obtain the accuracy on $D_{\mathrm{V}}$ motivates the goal for the \textit{Dark Energy Explorers} citizen science campaign.}
      \end{center}
      \label{fig:MainPlot}
\end{figure*}

The accuracy on the cosmological constraints coming from HETDEX is primarily determined by the number of LAE sources, the false positives due to noise (FP) and the \OII contamination. We have optimized our detection and classification algorithm \citep{Davis2022,Gebhardt21,erincooper, Farrow2021, Leung2017} to reach specifications on each of these parameters. This paper presents how we utilize the human eye’s ability for pattern recognition to further improve on the project's algorithm and possibly push into new regimes, such as lower signal-to-noise and poor emission-line fits.

Citizen science is a collaboration between scientists and the public to reach a larger science goal. Aiming to utilize the human eye’s ability for pattern  recognition we created a citizen science project, \textit{Dark Energy Explorers}\footnote{\url{https://www.zooniverse.org/projects/erinmc/dark-energy-explorers}}, to improve HETDEX data products. 

Other large surveys have used contributions from citizen science as part of the analysis pipeline e.g., Gravity Spy for LIGO to improve noise rejection \citep{GravitySpy}, TESS's Planet Hunters  \citep{PlanetHunters} and Catalina Outer Solar System Survey's \citep{Catalina} evaluation of candidates to improve completeness. \textit{Galaxy Zoo} is a well-established project to classify galaxies from SDSS and it has set the stage for large-survey citizen science \citep{SDSSGalaxyZoo2008, SDSSotherGalaxyzoo2008}. Beginning as a simple website, \textit{Galaxy Zoo} has now expanded and created a host platform, \textit{Zooniverse}\footnote{\url{https://www.zooniverse.org/}}, which is now the world’s largest citizen science platform. \textit{Zooniverse} is home to dozens of citizen science projects in various disciplines that have led to hundreds of publications. \textit{Dark Energy Explorers} is one of these programs. \textit{Zooniverse's} mission is “to enable research that would not be possible, or practical, otherwise." Building off of this mission, \textit{Dark Energy Explorers} aims to improve the accuracy of HETDEX, while simultaneously allowing participants the first look at astronomical sources by teaching them to classify millions of sources from the Hobby-Eberly Telescope. 

HETDEX has no pre-selection of sources, which is quite different from other large-scale surveys. We tile the sky over 540 square degrees with a fill factor of 1/4.5 using 74 integral field units. These units feed 156 spectrographs covering a wavelength range of 350-550nm, with a resolving power that ranges from 750-950. The software then searches through every spectral and spatial resolution element for emission lines, including those from LAEs. Over its lifetime, HETDEX will acquire about one billion spectra and one trillion resolution elements (spatial and spectral). From these trillion resolution elements we expect to find about 1.3 million LAEs, 0.92 million \OII emitters, and many stars, meteors, asteroids etc. The reduction from one trillion resolution elements to the one million LAEs is dependent on the signal-to-noise ($S/N$) limit adopted by the experiment. At lower $S/N$, more emission lines are detected, but at the expense of more false positives due to artifacts \citep{erincooper}. HETDEX is exploring ways to use all trillion resolution elements, which requires accurate control of pixel level defects.

One fundamental challenge of HETDEX involves sifting through the data and distinguishing the LAEs from the other detections, including false positives, \OII emitting galaxies, meteor trails and other line-like features. HETDEX has optimized its algorithms to distinguish the various sources, but there are advantages that the human eye can provide that will improve the survey. While it is impractical for a small research team to visually vet many millions of sources, we realize that, if possible, such vetting would enable significant improvements towards the measurements of $H(z)$ and $D_A(z)$. Our goal is to use citizen science to classify millions of HETDEX sources and help keep the contamination rate low.

Visual vetting has two important aspects. First, we use citizen science to help identify sources caused by non-Gaussian noise, therefore reducing the false-positive rate and generating a cleaner training set for machine learning. Second, we are able to explore regimes that are more difficult to classify algorithmically, such as that for detections with low signal-to-noise and/or higher chi-squared from a single emission-line fit (see Section~\ref{subsec:interface}).  For example, active galactic nuclei (AGN) have a variety of emission profiles that are easy for the human eye to distinguish but can be difficult for a general algorithm that uses a single unresolved emission line. By including these additional sources, we improve our accuracy on the expansion rates. The current results suggest that we can increase the HETDEX LAE sample by up to $50\%$, which yields a $\sim 20\%$ improvement on the distance estimates (see Figure \ref{fig:MainPlot}). Our goal is to combine the visual vetting with machine learning, and the first step towards doing this is to create labels for training the machine learning model. Creating these labels to train the machine creates a rich opportunity for \textit{Dark Energy Explorers} and HETDEX.

This paper discusses how we use \textit{Dark Energy Explorers} to help HETDEX reach specifications. In Section~\ref{sec:improvements}, we discuss how to improve the accuracy of the distance measures using the collected data from \textit{Dark Energy Explorers}. Section~\ref{sec:explorers} focuses on \textit{Dark Energy Explorers} interface and how we train the public to become HETDEX astronomers. Section~\ref{sec:deeresults} presents initial results of \textit{Dark Energy Explorers}. Section~\ref{sec:machinelearning} shows how we incorporate the \textit{Dark Energy Explorers} in the HETDEX database using machine learning, and Section~\ref{sec:discussion} will discuss conclusions.

\section{Improving the Accuracy of the Distance Measures from HETDEX}
\label{sec:improvements}

HETDEX will obtain about one billion spectra and one trillion resolution elements over the lifetime of the survey \citep{Gebhardt21, erincooper}. Once HETDEX reaches completion, we expect to have over one million redshifts of distant LAEs between $1.88 < z < 3.52$, and over one million redshifts for nearby \OII galaxies with $z < 0.5$. These galaxies are what we use for the cosmological analysis. The requirements are a false positive rate  $< 10\%$ due to noise, contamination due to \OII emitters $< 2\%$, and the total number of LAEs of over one million. As shown in \citet{erincooper} and \citet{Davis2022}, we reach the specifications with little margin for error. Our goal here is to push to lower signal-to-noise and higher chi-squared, measured against a Gaussian fit. This will increase the number of LAEs and still maintain the low false positive and contamination rates. In order to keep these rates this low, we employ visual vetting and machine learning. 

The HETDEX HDR2 contains $\sim 50,000$ objects with well-fit emission-line profiles (S/N $> 5.5$ and $\chi^2 < 3$) that are classified as LAEs \citep{erincooper}. This catalog is from contiguous fiber spectra coverage of 25 deg$^2$ of the  sky. There are undoubtedly many sources outside these cuts; for example AGN with broad emission lines will deviate from a single-line fit causing the chi-square value to be high. We expect a robust visual vetting and machine learning campaign to extract an additional $\sim 10\%$ of the catalog, while keeping the false positive and contamination rate low. 

Figure \ref{fig:MainPlot} quantifies the trade between S/N cut, chi-squared cut, false positive rate, and contamination rate for the primary cosmological parameter $D_{\mathrm{V}}$. For this figure, we need to assume the LAE galaxy bias, since the combined distance accuracy improves linearly with bias (higher bias values provide better accuracy). We are in the process of measuring this bias accurately (Farrow et al. in preparation), and for this figure we assume $b_{LAE}=2.0$. For the properties we have control over (S/N, FP, chi squared, \OII contamination), we base these rates from analyses by \citet{Davis2022} and \citet{erincooper}. As demonstrated in Figure \ref{fig:MainPlot} the accuracy of $D_{\mathrm{V}}$ depends on all three factors. The black horizontal line represents the target goal for the entire HETDEX survey. The primary relation is that $D_{\mathrm{V}}$ goes as the inverse square root of the number of LAEs, and this is given by the gray line assuming no corrections for false positive rate or \OII contamination. The vertical lines represent the different S/N cuts as given in the legend. The distance accuracy is degraded by the contribution from the false positives (the blue line), and then further including \OII contamination (the red line). The uncertainty on $D_{\mathrm{V}}$ scales linearly with the false positive rate, as the false positives affect both the number of true LAEs and add in white noise. The uncertainty on $D_{\mathrm{V}}$ scales quadratically with the \OII contamination, as this affects the number of LAEs and imposes clustering power from nearby galaxies onto the LAE power spectrum. \cite{grasshorngeb} show that an upper limit of 2\% on \OII contamination does not impact the scientific requirements significantly. Within Figure \ref{fig:MainPlot} we assume to have a \OII contamination rate of 0.013, based on work of \cite{Davis2022}. The false positive rate is a result of measuring a confirmation rate of LAE's and therefore our FP rate will be less than the confirmation rate. Initial FP rate estimates are given in \cite{Gebhardt21} and \cite{erincooper} to be $<$0.24 at  $S/N = 4.8$, $<$0.19 at  $S/N = 5$, $<$0.05 at  $S/N = 5.5$ and $<$0.01 at  $S/N >6$, providing upper limits. 

The intersection of the gray, blue, and red curves with the black horizontal dashed line provides the uncertainty on $D_{\mathrm{V}}$. The red line has the largest uncertainty on $D_{\mathrm{V}}$, since it accounts for contamination and false positives, and is what we use for the final prediction. One can then read off the S/N cut that is required to obtain the desired accuracy on $D_{\mathrm{V}}$, motivating our goal for the \textit{Dark Energy Explorers} citizen science campaign.

Given that the primary corrections are due to false positives and \OII contamination, we focus our citizen science campaign on those issues. As described below, the false positive rate is a visual distinction between real and fake sources, and the \OII contamination is a distinction between distant and nearby galaxies.

\begin{figure*}[ptb]
    \begin{center}
         \includegraphics[scale=0.8]{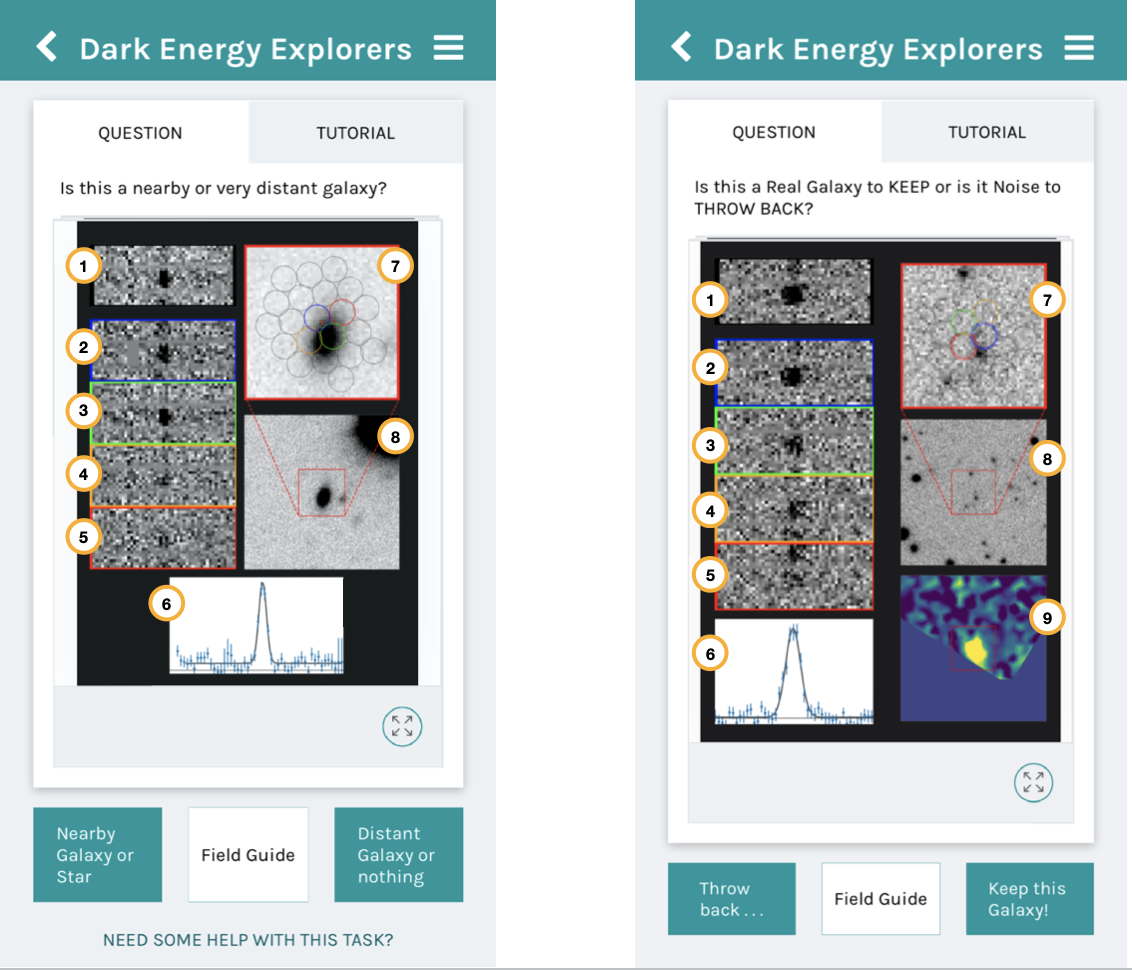}
  \caption{Above shows the mobile version of \textit{Dark Energy Explorers}. In either workflow (`Nearby vs.\ Distant Galaxies' on the left, `Fishing for Signal in a Sea of Noise' on the right), the participants are shown various visualizations of a HETDEX detection from the Hobby-Eberly Telescope (HET) and given a binary choice to classify the astronomical source. Access to the tutorial is shown in the top right corner and the field guide in the bottom center. Image Summary: (1) - (5) 2D Fiber Cutouts, (6) 1D Emission Line Fit, (7) Imaging Postage Stamp with Fiber Positions (zoomed in, $9\arcsec \times 9\arcsec$), (8) Imaging Postage Stamp (zoomed out, $30\arcsec \times 30\arcsec$) (9) \textit{Lineflux Map}. Image (9) is only included in the `Fishing for Signal in a Sea of Noise' workflow.}
  \label{fig:labeled8mobile}
      \end{center}
\end{figure*}

\section{Developing Dark Energy Explorers}
\label{sec:explorers}

Through the end of 2022, HETDEX data collection is $50\%$ complete and analysis from the first internal data release in 2019 showed the primary need is for careful visual vetting of the source catalog \citep{Gebhardt21}. To solve this problem, we designed, created, and launched the worldwide citizen science project, \textit{Dark Energy Explorers}, in late February 2021.

The goal of \textit{Dark Energy Explorers} is to solve the problem of having an infeasible amount of data to classify with a small in-house team. Initial visual vetting by HETDEX collaboration members provided source classifications, but took too much time to make classification of the full dataset feasible. Establishing a project on \textit{Zooniverse} facilitated participation from thousands of volunteers across the world drawn from the platform's millions of registered volunteers where the project is accessible on any smartphone, tablet, or desktop computer with internet access. In addition to aiding in volunteer recruitment, the Zooniverse team has provided support from project creation, to launch, and beyond. In addition to reaching a larger science goal, the project tutorial (discussed more in Section~\ref{subsec:tutorial}) is intended to allow anyone to participate, even those without a science background. 

We constructed two different workflows for \textit{Dark Energy Explorers}: “Nearby vs.\ Distant Galaxies" and “Fishing for Signal in a Sea of Noise." The first workflow aims to differentiate between \OII emitting objects at $z<0.45$  (“Nearby Galaxies") from $1.88<z<3.52$ LAEs (“Distant Galaxies"). The `Nearby vs.\ Distant" galaxies workflow addresses \OII contamination and helps us optimize our discrimination algorithm. Therefore, the target objects all have $S/N > 5$ and include candidate LAEs, AGN, \OII galaxies, and stars.  
The second, more recent workflow, `Fishing for Signal in a Sea of Noise', addresses the false positives. As the false positives caused by noise, they are  more difficult to identify. This workflow therefore hones in on classifying the $S/N > 6$ LAE candidates. 
This is the subset that is addressed in detail with machine learning efforts in this paper and lower signal to noise regimes will be explored in future work. 

\subsection{Tutorial and Field Guide}
\label{subsec:tutorial}

\textit{Dark Energy Explorers} is accessed via the \textit{Zooniverse} app or website. Participants can create an account to save classifications and data or choose to participate anonymously. The \textit{Dark Energy Explorers} project can be accessed directly via its URL or via selection from the Zooniverse projects page where it is listed under the “Space" or “Physics" categories. Volunteers can then choose a workflow from the project landing page, either “Nearby vs.\ Distant Galaxies" or “Fishing for Signal in a Sea of Noise." After choosing the workflow, the participants walk through a tutorial on how to classify the HETDEX data (discussed more in Section~\ref{subsec:interface} and Figure \ref{fig:labeled8mobile}).

The foundation of \textit{Dark Energy Explorers} is the easy-to-understand tutorial. The tutorial trains members of the public to become amateur HETDEX astronomers without any prior astronomy or general science background, by simplifying the classification process into digestible, jargon-free tasks. The participants read a tutorial and are provided a number of criteria to allow them to choose between two binary options in the workflows. We opted to limit the classifications to a binary choice to maximize classifying speed. For example, on a mobile device, which offers the swipe-left/swipe-right classifying option, a \textit{Dark Energy Explorer} can classify twenty sources in less than a minute. This ensures that users have the ease of the swiping feature on the mobile device with simple choices, avoiding multiple or nested selections. 

\begin{figure*}[ptb]
    \begin{center}
         \includegraphics[scale=0.6]{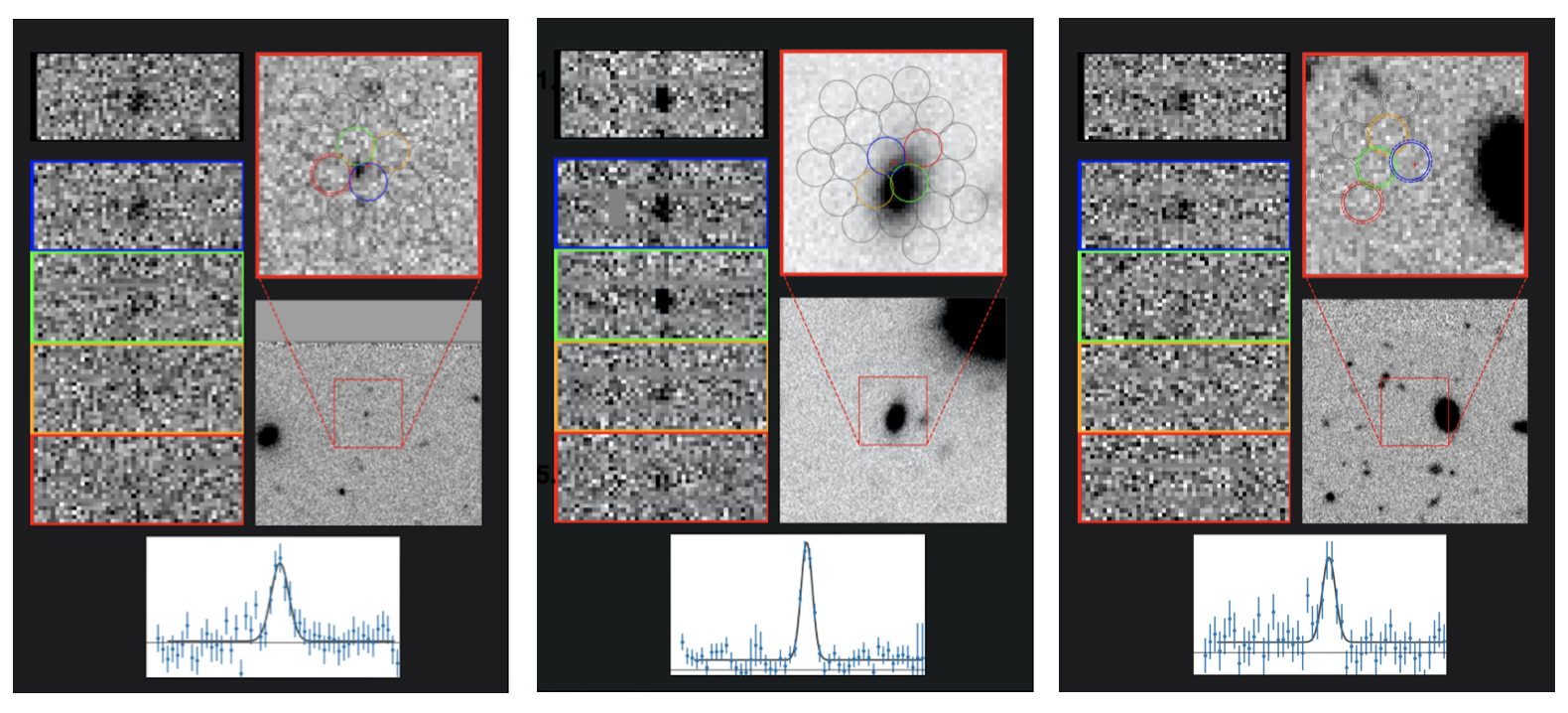}
  \caption{Above we have examples of the “mini's" from \textit{Dark Energy Explorers} “Nearby vs.\ Distant" workflow. From left to right: A distant Lyman-$\alpha$ emitting galaxy, a nearby \OII emitter, and a tricky case that might need more information.}
  \label{fig:NearbyDistantTossup}
      \end{center}
\end{figure*}

The primary criteria users have to consider for the “Nearby vs.\ Distant" workflow are:

\begin{enumerate}
    \item The relative size of the object, 
    \item The strength of the emission line, and
    \item The appearance of the emission line in at least one or more of the fiber spectrum. 
\end{enumerate}

For example, a nearby galaxy often appears as a large, bright resolved source in the images, while distant galaxies are generally small and faint. See Figure \ref{fig:NearbyDistantTossup} for a comparison. 

\begin{figure*}[ptb]
    \begin{center}
         \includegraphics[scale=0.6]{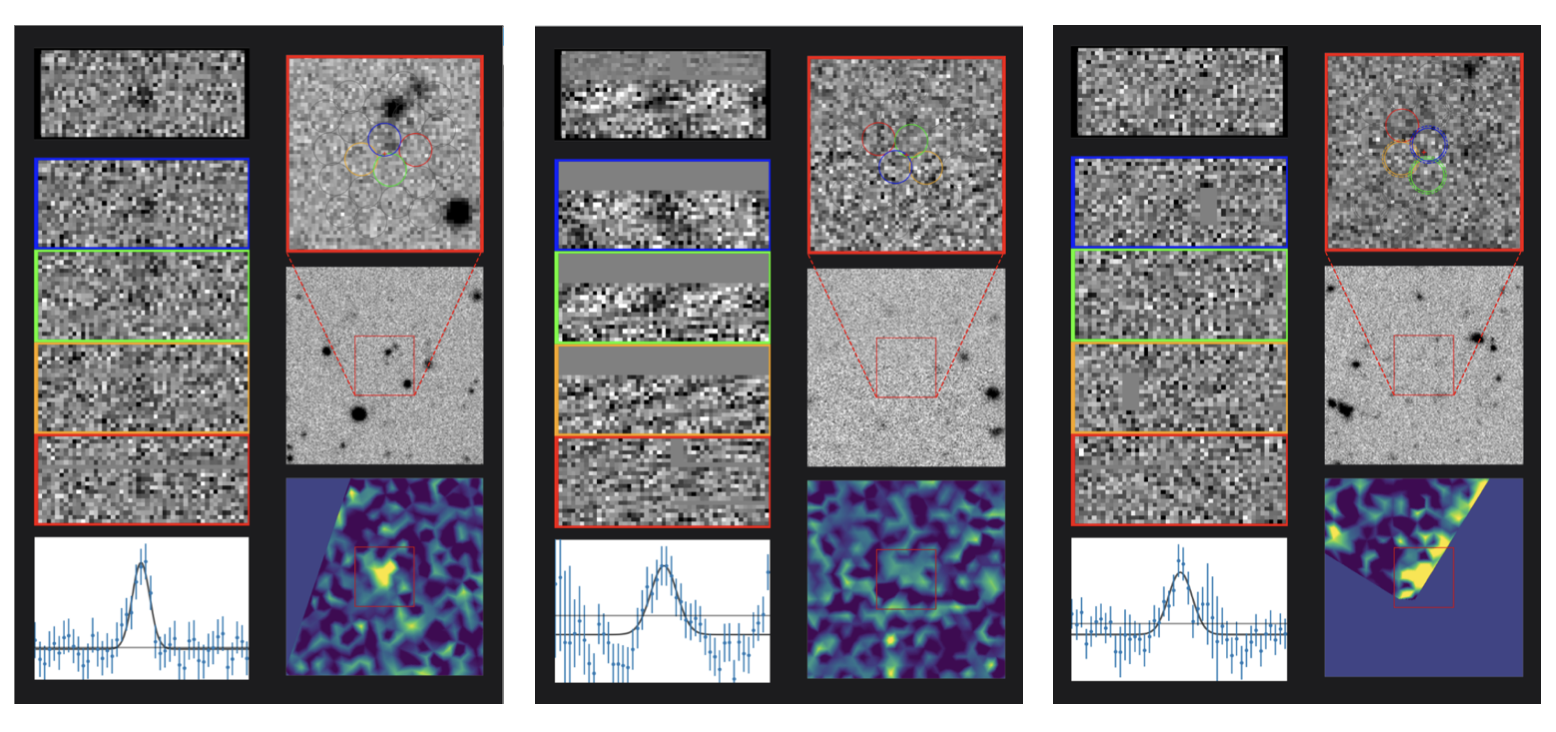}
  \caption{Above we have examples of the ``mini's" from \textit{Dark Energy Explorers} ``Fishing for Signal in a Sea of Noise" workflow.  From left to right: Keep (real galaxy/emission line), Throwback (bad detection), and a tricky case that might need more information.}
  \label{fig:KeepThrowbackTossup}
      \end{center}
\end{figure*}

The primary criteria users have to consider for the “Fishing for Signal in a Sea of Noise" workflow are:

\begin{enumerate}
    \item The quality of the data collected,  
    \item The strength of the emission line, and
    \item The appearance of the emission line in at least one or more of the fiber spectrum.
\end{enumerate}
See Figure \ref{fig:KeepThrowbackTossup} for a comparison. In addition to the tutorial, participants have a short easy-access field guide that is available to them on the main classification page. This guide gives a quick reminder of the selection criteria for either workflow. Figure \ref{fig:labeled8mobile} shows the mobile version of \textit{Dark Energy Explorers} where participants have access to both the field guide and an option to go back to the entire tutorial at anytime during classification. 

\subsection{Interface and Data Inputs} 
\label{subsec:interface}

We have combined multiple images and spectral data to create what we call a “mini" image. We refer to them as “mini's" because they are a greatly reduced and compact representation of the more complete and complex Emission Line eXplorer (ELiXer) \citep{Davis2021}. These images were designed to be compact and compatible with both a desktop or mobile device. Because we desired the workflow to be usable on mobile devices with a variety of phone resolutions and aspect ratios, we decided to keep the visualizations simple and compact. In addition, we wanted the site to be accessible to people without science backgrounds; therefore the images are free of jargon and numbers which could easily distract or turn away participants.  

The “mini's" consist of panels (1) - (8) for the “Nearby vs.\ Distant Galaxies" workflow and images (1) - (9) for the “Fishing for Signal in a Sea of Noise" workflow. See Figure \ref{fig:labeled8mobile}. These images are defined in detail here:

(1) - (5) \textit{2D Fiber Cutouts}: Five cutouts within ±40 Å of the detection line center in the spectral direction and ±1 fiber across the detector. The spectral images are sky subtracted. Image (1) (highlighted in black) is the weighted sum of all contributing fibers. The rows below, images (2) - (5) (blue, green, orange, red) are the four fibers ordered by distance from the source position. 

(6) - \textit{1D Line Fit}: The resultant 1D spectrum and the emission line fit.

(7) \textit{Postage Stamp with Fiber Positions}: The footprint of all fibers contributing to the detection plotted over deep ground-based imaging of a $9\arcsec \times 9\arcsec$ region centered at the spatial position of the maximum SNR of the emission-line. For internal classification purposes, a number of imaging catalogs are available as described in detail in \citet{Davis2022}. However, for the current workflows, we only display ancillary  $r$-band imaging obtained from Hyper Suprime-Cam on the Subaru Telescope by the HSC Subaru Strategic Program (HSC-SSP) \citep{HSC-SSP} and the HSC HETDEX Survey (HSC-DEX) \citep{Davis2022}. This limits confusion created by heterogeneous image quality and flux sensitivity. The four colored fibers match the colors outlined to the left in images (1) - (5). Fibers with a dashed outer ring are at the edge of the detector. The PSF weighted center of the detection is marked with a red cross.

(8) - \textit{Imaging Postage Stamp}: A postage stamp cutout of the from ancillary imaging data of a $30\arcsec \times 30\arcsec$ region centered at the spatial position of the emission line. This is a zoomed out version of image (7) to show possible catalog counterparts or nearby sources.

(9) - \textit{Lineflux map}: The wavelength collapsed flux intensity map over a $\pm3\sigma$ region from the emission line center. The lower section of this particular map is blank as the region happens to fall off the edge of the detector. For more detail about how these minis are generated, see \cite{Gebhardt21} and \cite{Davis2022}.

These telescope images are of a detection on a random area of sky, centered on an object (i.e. galaxy, star, AGN) chosen at random from the sample database. \textit{Dark Energy Explorers} avoids using numbers and astronomical jargon within the tutorial and field guide, making an approachable way to train anyone to become a volunteer for HETDEX.

\subsection{Education and Engagement}
\label{subsec:education}

While the main focus of \textit{Dark Energy Explorers} is for science advancement of HETDEX, other primary goals have been for education, outreach and public engagement. To do this, we have implemented a feature to educate participants while they conduct galaxy classifications. Other projects, such as Galaxy Zoo, have used this method to provide continued engagement and longer participation in the projects (\citealt{SDSSGalaxyZoo2008, SDSSotherGalaxyzoo2008}). Learning from the Galaxy Zoo project, \textit{Dark Energy Explorers} aims to provide ways for participants to interact with each other and the HETDEX team to learn more of the science \citep{pratherWallace}. Following these methods we have incorporated a similar objective for our consistent users of Dark Energy Explorers. For every 50 sources an individual classifies, they receive a pop-up message linking them to learn more about the science behind HETDEX, dark energy, and general astronomy. This “level-up” incentive provides continued engagement with the project while educating the public about astrophysics topics and empowering them as scientists. 

In effort to engage with our participants and address questions they may have, \textit{Dark Energy Explorers} has a ``Talk" board that allows participants to post comments and/or questions. These ``Talk" boards are monitored by the HETDEX team and allows the participants to engage with the scientists or ask frequent questions. So far, we have had over 500 participants engage with the discussion boards and post over 5000 comments and images. The discussion boards provide a space for participants to see many examples of classified sources for each workflow. The HETDEX team member can comment back to confirm or deny the classifications to address these questions and other participants can then see the discussion.

\section{Dark Energy Explorers Results}
\label{sec:deeresults}

\subsection{Classification and User statistics} 

\textit{Dark Energy Explorers} launched in February 2021 and since then has collected roughly three and a half million classifications by 11,000 volunteers in over 85 different countries around the world. See Table \ref{tab:1} for workflow classifications and selection from the HETDEX data releases. This is a result of over three million classifications within the ``Nearby vs. Distant" workflow. These classifications amount to $\sim$209,000 sources classified as LAE's or \OII galaxies. The ``Nearby vs. Distant workflow was originally launched workflow and thus has more classifications in total. The ``Fishing for Signal in a Sea of Noise" was launched in October of 2021, and has a total of 140,000 classifications resulting in  $\sim$14,000 completed subjects which are the focus of the machine learning efforts (see Section \ref{sec:machinelearning}).

Since launch, roughly $60\%$ of our users have classified more than 50 objects and have received the pop-up (see Section \ref{subsec:education}). Of those users two-thirds have had continued engagement with the project by completing over 100 classifications. In other words, over $40\%$ of total participants have made at least 100 classifications. 

With visual vetting we acknowledge human bias, and also work to build in reliability in human classification. In order to have a robust entry for each source, we must average over some number of participants.  The number of classifications selected per subject is a balance between classification efficiency and ensuring incorrect classifications are mitigated by statistics.  We found that 10 classifications yield values that are just as consistent as the larger sample sizes and allows many more objects to be classified. The benefit of the binary classification is that it is very straightforward to obtain a classification statistic from a workflow by simply taking a median or average of the individual classifications. For this work, the ten binary human classifications are averaged to build confidence in an accurate cataloguing \citep{Santos2021}. The final product yields a \textit{Dark Energy Explorers} probability (DEE probability) that the source is a distant galaxy or real signal by visual vetting, according to each workflow, respectively. 

We note that the standard deviation of the $N$ classifications is also of potential use as it indicates objects where there is strong disagreement on the subject amongst the citizen scientists. Internal HETDEX team members follow up these special cases by eye. The \textit{Dark Energy Explorers} measurements thus reduce the number of sources the small internal HETDEX team must visually investigate and offers an effective method to identify rare, unique sources within the HETDEX survey. Out of 140,000 total classifications in “Fishing for Signal in a Sea of Noise”, 0\% have a standard deviation greater than 0.5.

\subsection{Nearby vs Distant Results}

The citizen scientist data from the “Nearby vs.\  Distant" workflow is working to validate current research by comparing the visual DEE probability to the model probability \citep{Davis2022}. We had over three million classifications by the \textit{Dark Energy Explorers} which resulted in 209,000 sources classified, since each is viewed by a minimum of 10 different participants. Once we calculate the DEE Probability we can compare the \OII and LAE classifications. Where they disagree allows us to delve into those sources and determine where the disagreement stems. In some cases, the classification from the \textit{Dark Energy Explorers} participants are correct and we use these instances to improve our model \citep{Davis2022}. The \textit{Dark Energy Explorers} classifications match the model, EliXer, with more than 92\% agreement. The success of the  \textit{Dark Energy Explorers} significantly reduced the time spent by the HETDEX team manually inspecting sources \citep{Davis2022}. This is due to the fact that we used the ``Nearby vs.\ Distant" workflow provide a way to incorporate visual vetting of the problem areas to tune our algorithm. With theses effort from \textit{Dark Energy Explorers} and we now can rely on \cite{Davis2022} for the \OII contamination rate (See Figure \ref{fig:MainPlot}). Because of the work of the citizen scientists to confirm the \OII emitting sources we have retired this workflow and focus on the machine learning efforts with ``Fishing for Signal in a Sea of Noise."

\begin{table*}[t]
\caption{\textit{{Dark Energy Explorers workflows}}}
\centering
\begin{tabular}{ |c|c|c|c|c|c|c|}
 \hline
 \hline
Workflow Name & Launch & Parent  & S/N cut & Down & Sources & Total  \\ 
              & date    & sample &  & Selection & complete & classifications \\ 
 \hline
 \hline
Nearby vs. Distant & Feb 2021  & 500,000   & 5.5 & 209,000 & 114,000 LAE & 3.1 M\\ 
&  & & &  & 95,000 [OII] & \\ 
\hline
Fishing for Signal in a Sea of Noise & Oct 2021 & 700,000 & 6 & 60,000 & $14,000^a$ & 140,000\\ 
 \hline
  \end{tabular}
\footnotesize{\\ $^a$ Ongoing data collection\\}
  \label{tab:1}
\end{table*}

\subsection{Fishing for Signal in a Sea of Noise Results}


The raw data products of HETDEX contain many pixel-level events, such as charge traps, hot pixels, and time-variable changes in calibration. Since these imperfections are often hard for algorithms to identify \citep{Gebhardt21}, most have to be found manually \citep{erincooper}. The human eye is good at identifying these features, particularly at high $S/N$, since they tend to have correlated residuals or create obvious features in the 2D charge coupled detector (CCD) imaging. Thus the main goal for \textit{Dark Energy Explorers} is to generate an additional removal of false sources from the catalog. For this paper we focus on the false positives, so that a DEE probability of 1.0 means an object is real, and a probability of 0.0 identifies a false detection. Thus, DEE probability is only in reference to the `Fishing for Signal in a sea of Noise" workflow. 

\begin{figure*}[ptb]
    \begin{center}
         \includegraphics[scale=0.6]{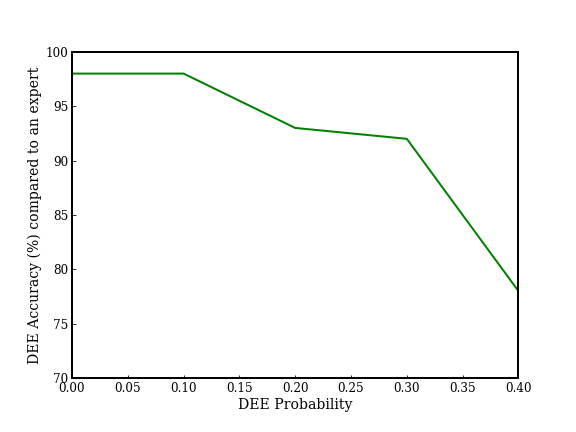}
  \caption{{Above shows the percent accuracy of the \textit{Dark Energy Explorers} participants as compared to a HETDEX expert for the `Fishing for Signal in a Sea of Noise" workflow. This accuracy demonstrates how well the \textit{Dark Energy Explorers} participants perform compared to a HETDEX team member. The DEE probability is the average of the 10 classifications for each source and represents the probability that a source is a false positive -- 0 -- or a real detection -- 1. This figure and the analysis in this paper focus on identifying the false positives or DEE probabilities $<$ 0.4.}} 
  \label{fig:DEEAccuracy}
      \end{center}
\end{figure*}

Since the end of 2022, we have had over 140,000 classification for the ``Fishing for Signal in a Sea of Noise" workflow, resulting in 14,000 sources out of a total of 60,000 input. Figure \ref{fig:DEEAccuracy} shows the how well the participants do (i.e., the accuracy of the \textit{Dark Energy Explorers}) compared to a HETDEX team member against the DEE probability. For the figure, the data have been binned to intervals of 0.1 in probability with 200 sources per bin.  This analysis determines the cut in DEE probability where we accurately remove false detections without removing too many real sources. At this point, the main interest for the \textit{Dark Energy Explorers} is to remove the obvious false positives. Again, when calculating the DEE probability, a false source and a real source corresponds to 0 and 1, respectively. Thus, we focus on the accuracy of the \textit{Dark Energy Explorers} classifications when the source is likely to be an artifact (i.e., having a probability less than 0.4).  The plot indicates that \textit{Dark Energy Explorers} most often agree with the HETDEX astronomers when the probability of the object being a real source is low. Below we describe the strategy for aggregating the results from the \textit{Dark Energy Explorers} probabilities and combine their classifications of artifacts with machine learning efforts.

\begin{figure*}[ptb]
    \begin{center}
    \includegraphics[scale=0.37]{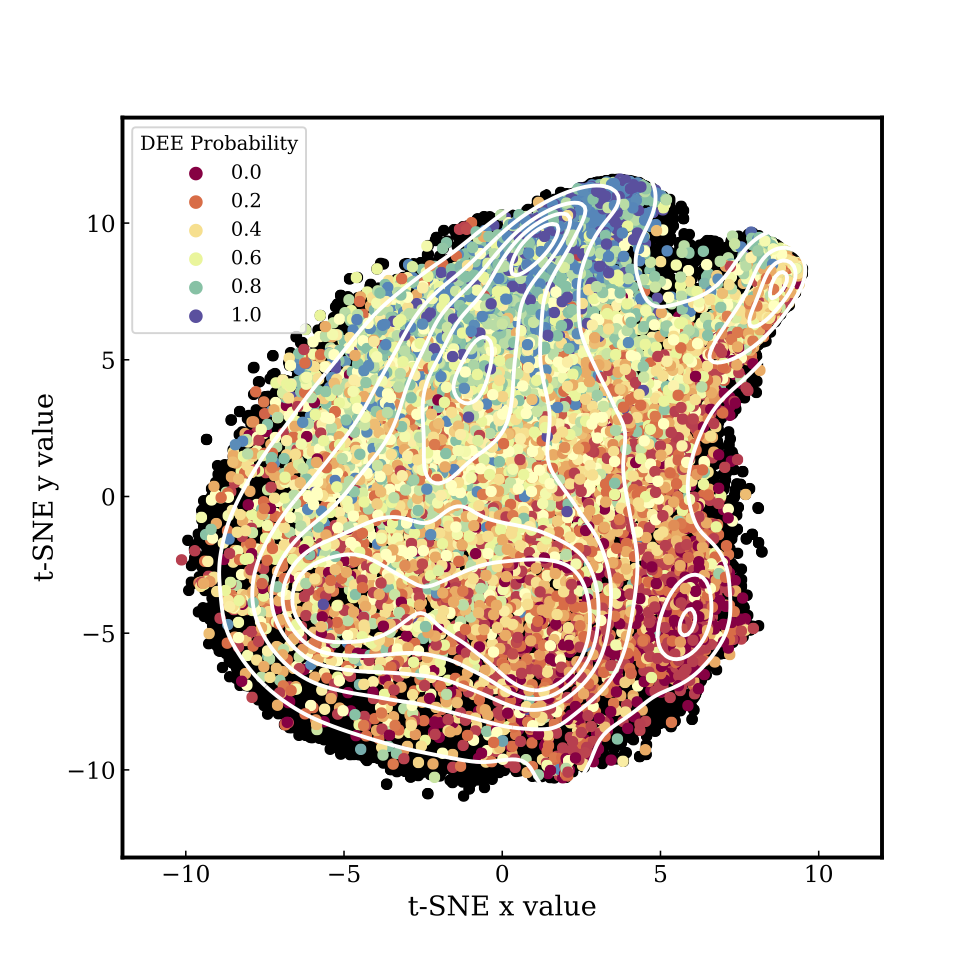}
  \caption{{Above shows a visual representation of the t-SNE machine learning algorithm. The colored points are labels created from the \textit{Dark Energy Explorers} probabilities ($\sim$ 14,000) which are overplotted on the black points which represent the total number of sources with S/N $> 5.5$ within the source catalog ($\sim$ 120,000 sources). The contours of source density within the t-SNE space are shown in white.}}
  \label{fig:DEEmachinelearncontour}
      \end{center}
\end{figure*}

\section{Incorporating the Dark Energy Explorers Results with Machine Learning}
\label{sec:machinelearning}

The primary goal is to use the \textit{Dark Energy Explorers} to improve the HETDEX catalog. Since HETDEX is a multiyear project that uses a large number of detectors, there are a significant number of instrumental and calibration issues that can produce features that mimic real sources. A secondary goal is to use the classifications as a training set for unsupervised learning efforts. We analyze the citizen science outputs with both of these in mind. 

To further ensure removal of the false positives from the data, we explore here including machine learning into the overall classifications. There are a few aspects worth considering. First, we can use the DEE classification as labels for machine learning. Second, we can use machine learning to help inform the \textit{Dark Energy Explorers} training. Third, we can combine the \textit{Dark Energy Explorers} classifications with machine learning to provide stronger leverage on the final classifications. For example, unsupervised learning combined with the \textit{Dark Energy Explorers} probabilities would allow us to apply cuts to the full sample of sources as an additional technique to remove false detections from the catalog.

We have about 140,000 classifications by participants since each source is inspected by a minimum of 10 individuals. Of those 14,000 categorized detections, $\sim$ 2,000 have a DEE probability below 0.1 which are considered the mostly likely to be false positives. Furthermore, $\sim$ 4,000 sources have a DEE probability less than or equal to 0.3. From Figure \ref{fig:DEEAccuracy} the agreement of identifying false positives (DEE probability $\le 0.3$) by the \textit{Dark Energy Explorers} stays above 92\%. When the DEE probability falls between $0.3 - 0.7$, the accuracy also falls. These are the tricky cases as demonstrated in Figure \ref{fig:KeepThrowbackTossup} and we must turn to further information (such as EliXer) and the HETDEX team to gather more information to classify the source.

For the purpose of reducing the false positives within HETDEX we will focus our analysis on the false positives. We use unsupervised learning to map \textit{Dark Energy Explorers'} probability values to spectral characteristics. We do this via t-distributed stochastic neighbor embedding, t-SNE \citep{vandermaaten08}, from the {\tt scikit-learn} Python package \citep{scikitlearnPedregosa11}. 

We analyzed t-SNE inputs of both the full spectra and with a $\pm$ 50~\AA\ region centered around the detected emission feature. t-SNE then takes the high-dimensional data set and produces a representation of those points in a lower-dimensional space. Here we will reduce the data to two-dimensions and therefore the n-components are set to 2. The next parameter we set is the perplexity. Following the most recent research, the perplexity is shown to have optimal results when set as the square root of the number of data points or fixed at 50 for large data sets \citep{Maaten2015}.

Datasets of high dimensionality, like our spectral data (1036 dimensions for the full spectra) often presents challenges when analyzing (\citealt{vandermaaten08}; Valentina Tardugno Poleo, in prep). For this analysis we use only a small wavelength region ($\pm$ 50\AA) around the emission line. We do this cut in order to focus on the line itself and keep it centered for the t-SNE visualization.  We have tried using the full spectrum, and in this case t-SNE tends to separate based on redshift. Since we are primarily focused on real emission or false positives, we currently remove the redshift information. Future work will continue to explore using the full spectrum. For example, applying a PCA analysis or an autoencoder might help in removing the redshift from the visualization, and help to further discriminate sources. Therefore, reducing the dimensions of the dataset, from 1036 for the full spectra down to 100 for the cut-out around the emission line, allows us to avoid the complications and simplifies the data visualization. Making this dimension reduction gave results with better discrimination and those results are discussed here.


In Figure \ref{fig:DEEmachinelearncontour} we show the combination of the \textit{Dark Energy Explorers} participants with the t-SNE machine learning technique, which results in the highest removal of fake sources and the lowest percentage of removal of real detections. Figure \ref{fig:DEEmachinelearncontour} shows a visual representation of the t-SNE machine learning algorithm. The colored points are labels created from the \textit{Dark Energy Explorers} probabilities ($\sim$ 14,000) and the black points are the total number of sources with S/N $> 5.5$ within the source catalog ($\sim$ 120,000 sources). The contour of source density within the t-SNE space is overplotted in white. These results demonstrate how the combination of citizen science visual vetting and t-SNE can yield the more consistent results. 

Going through all the sources that have a \textit{Dark Energy Explorers} probability of below 0.1 we find a result of $98\% $ agreement, as seen in Figure \ref{fig:DEEAccuracy}. All of these sources are reviewed by members of the HETDEX team and then compared to the \textit{Dark Energy Explorers} probability to determine this accuracy.

The variety of problems we see in the significant number of detectors and resolution elements make it hard for us to quantify all categories of false positives, which is why we first employ members of the HETDEX team. The t-SNE visualization will eventually be combined with a supervised learning algorithm where we can group regions, providing a well-trained classification scheme. Our goal for this paper is to determine the false positives and remove them from the sample with the lowest amount of real detections being removed. Our next goal is to use these vetted false positives to enhance the classifications and the selection configuration.

We find that the highest accuracy of both \textit{Dark Energy Explorers} and t-SNE is within the region of t-SNE $x$ values $> 3$ and t-SNE $y$ values $< -1$ using the visualization from Figure \ref{fig:DEEmachinelearncontour}. The selected t-SNE region also yields the lowest removal of real detections with none of sources in this region having a DEE probability of $>$ 0.9. In other words, using this combination of methods no real sources, determined by visual vetting, are removed in the process. In addition, only 0.08\% of sources in this region have a DEE probability of $>$ 0.6, which are real detections according to the  \textit{Dark Energy Explorers}. Since we have visual confirmation that professional astronomers believe the sources to be real we will use the DEE probability to assure that these sources stay in the catalog, meaning that no real detections will be removed. This technique demonstrates how using the combination of machine learning and citizen science allows for isolated removal of false detections while the real detections remain in the catalog.

These results from \textit{Dark Energy Explorers} allow us to use t-SNE and expand on the full catalog of $\sim 120,000$ sources, which are the black points in Figure \ref{fig:DEEmachinelearncontour}. By applying the selected region of t-SNE space we can remove sources while assuring real detections will remain in the sample because we have visual confirmation from the \textit{Dark Energy Explorers}. This work results in the removal of 7,871 false detections from the catalog.

We can expand on this cleaning technique to see where other sources in the current catalog lie. This provides an efficient method to remove sources from the catalog without having to be completely vetted by humans. In conclusion, citizen science and machine learning when used synergistically can enhance science goals while creating a unique educational opportunity for the public.

\section{Conclusions}
\label{sec:discussion}

Since the end of 2022, \textit{Dark Energy Explorers} has collected roughly 4 million classifications by 11,000 volunteers in over 85 different countries around the world. An additional goal of this campaign is to have visual vetting by the \textit{Dark Energy Explorers} of 100\% of all sources down to S/N $> 4.8$ or the full HETDEX catalog which we expect to contain about three million sources. Repeating this process with all sources at a lower S/N will provide a robust classification using a combination of algorithms, data visualization and human vetting. While this number is significantly larger than what we have covered already, we will explore ways to engage a larger audience, to use a smaller number of individual classifications, and to incorporate targeted work flows.

So far, \textit{Dark Energy Explorers} has reduced our team’s visual vetting work, making this detailed level of research viable; it would have otherwise been impossible with our small team. This project is an innovative approach to studying dark energy and to classifying large data sets that require visual vetting. 

By using \textit{Dark Energy Explorers} we expect to improve constraints on the cosmological parameters by $\sim 10$ to $30\%$. For example, allowing us to go from S/N $> 5.2$ to S/N $> 4.8$  would result in a $30\%$ better accuracy on the cosmological parameter $D_{\mathrm{V}}$, following the motivation from Figure \ref{fig:MainPlot}. In addition to \textit{Dark Energy Explorers} being a uniquely powerful tool for science advancement, the project is increasing accessibility to science worldwide. 

\section{Acknowledgements}

The results of the Dark Energy Explorers would not be as robust and useful if not for the care and dedication by the volunteers. We are extremely grateful to their work. It is having a large impact and is motivational.

LH acknowledges support from NSF GRFP DGE 2137420 and NASA 21-CSSFP21-0009. KG acknowledges support from the NSF-2008793 and from NASA 21-CSSFP21-0009.

\textit{Dark Energy Explorers} is recognized as an official NASA Citizen Science partner. This publication uses data generated via the \href{https://www.zooniverse.org}{Zooniverse.org} platform, development of which is funded by generous support, including a Global Impact Award from Google, and by a grant from the Alfred P. Sloan Foundation. 

HETDEX is led by the University of Texas at Austin McDonald Observatory and Department of Astronomy with participation from the Ludwig-Maximilians-Universit\"at M\"unchen, Max-Planck-Institut f\"ur Extraterrestrische Physik (MPE), Leibniz-Institut f\"ur Astrophysik Potsdam (AIP), Texas A\&M University, The Pennsylvania State University, Institut f\"ur Astrophysik G\"ottingen, The University of Oxford, Max-Planck-Institut f\"ur Astrophysik (MPA), The University of Tokyo, and Missouri University of Science and Technology. In addition to Institutional support, HETDEX is funded by the National Science Foundation (grant AST-0926815), the State of Texas, the US Air Force (AFRL FA9451-04-2-0355), and generous support from private individuals and foundations.

The Hobby-Eberly Telescope (HET) is a joint project of the University of Texas at Austin, the Pennsylvania State University, Ludwig-Maximilians-Universit\"at M\"unchen, and Georg-August-Universit\"at G\"ottingen. The HET is named in honor of its principal benefactors, William P. Hobby and Robert E. Eberly. The Institute for Gravitation and the Cosmos is supported by the Eberly College of Science and the Office of the Senior Vice President for Research at the Pennsylvania State University.

The authors acknowledge the Texas Advanced Computing Center (TACC)  \footnote{\url{http://www.tacc.utexas.edu}} at The University of Texas at Austin for providing high performance computing, visualization, and storage resources that have contributed to the research results reported within this paper.

\bibliographystyle{aasjournal}
\bibliography{samples}

\begin{thebibliography}{}
\expandafter\ifx\csname natexlab\endcsname\relax\def\natexlab#1{#1}\fi
\providecommand{\url}[1]{\href{#1}{#1}}
\providecommand{\dodoi}[1]{doi:~\href{http://doi.org/#1}{\nolinkurl{#1}}}
\providecommand{\doeprint}[1]{\href{http://ascl.net/#1}{\nolinkurl{http://ascl.net/#1}}}
\providecommand{\doarXiv}[1]{\href{https://arxiv.org/abs/#1}{\nolinkurl{https://arxiv.org/abs/#1}}}

\bibitem[{{Aihara} {et~al.}(2018){Aihara}, {Arimoto}, {Armstrong}, {Arnouts},
  {Bahcall}, {Bickerton}, {Bosch}, {Bundy}, {Capak}, {Chan}, {Chiba}, {Coupon},
  {Egami}, {Enoki}, {Finet}, {Fujimori}, {Fujimoto}, {Furusawa}, {Furusawa},
  {Goto}, {Goulding}, {Greco}, {Greene}, {Gunn}, {Hamana}, {Harikane},
  {Hashimoto}, {Hattori}, {Hayashi}, {Hayashi}, {He{\l}miniak}, {Higuchi},
  {Hikage}, {Ho}, {Hsieh}, {Huang}, {Huang}, {Ikeda}, {Imanishi}, {Inoue},
  {Iwasawa}, {Iwata}, {Jaelani}, {Jian}, {Kamata}, {Karoji}, {Kashikawa},
  {Katayama}, {Kawanomoto}, {Kayo}, {Koda}, {Koike}, {Kojima}, {Komiyama},
  {Konno}, {Koshida}, {Koyama}, {Kusakabe}, {Leauthaud}, {Lee}, {Lin}, {Lin},
  {Lupton}, {Mandelbaum}, {Matsuoka}, {Medezinski}, {Mineo}, {Miyama},
  {Miyatake}, {Miyazaki}, {Momose}, {More}, {More}, {Moritani}, {Moriya},
  {Morokuma}, {Mukae}, {Murata}, {Murayama}, {Nagao}, {Nakata}, {Niida},
  {Niikura}, {Nishizawa}, {Obuchi}, {Oguri}, {Oishi}, {Okabe}, {Okamoto},
  {Okura}, {Ono}, {Onodera}, {Onoue}, {Osato}, {Ouchi}, {Price}, {Pyo}, {Sako},
  {Sawicki}, {Shibuya}, {Shimasaku}, {Shimono}, {Shirasaki}, {Silverman},
  {Simet}, {Speagle}, {Spergel}, {Strauss}, {Sugahara}, {Sugiyama}, {Suto},
  {Suyu}, {Suzuki}, {Tait}, {Takada}, {Takata}, {Tamura}, {Tanaka}, {Tanaka},
  {Tanaka}, {Tanaka}, {Terai}, {Terashima}, {Toba}, {Tominaga}, {Toshikawa},
  {Turner}, {Uchida}, {Uchiyama}, {Umetsu}, {Uraguchi}, {Urata}, {Usuda},
  {Utsumi}, {Wang}, {Wang}, {Wong}, {Yabe}, {Yamada}, {Yamanoi}, {Yasuda},
  {Yeh}, {Yonehara}, \& {Yuma}}]{HSC-SSP}
{Aihara}, H., {Arimoto}, N., {Armstrong}, R., {et~al.} 2018, \pasj, 70, S4,
  \dodoi{10.1093/pasj/psx066}

\bibitem[{Albrecht {et~al.}(2009)Albrecht, Amendola, Bernstein, Clowe,
  Eisenstein, Guzzo, Hirata, Huterer, Kirshner, Kolb, \& Nichol}]{kolb}
Albrecht, A., Amendola, L., Bernstein, G., {et~al.} 2009,
  \dodoi{10.48550/ARXIV.0901.0721}

\bibitem[{Alcock \& Paczyński(1979)}]{AlcockPacz}
Alcock, C., \& Paczyński, B. 1979, 281, 358, \dodoi{10.1038/281358a0}

\bibitem[{Bahaadini {et~al.}(2018)Bahaadini, Noroozi, Rohani, Coughlin, Zevin,
  Smith, Kalogera, \& Katsaggelos}]{GravitySpy}
Bahaadini, S., Noroozi, V., Rohani, N., {et~al.} 2018, Information Sciences,
  444, 172, \dodoi{https://doi.org/10.1016/j.ins.2018.02.068}

\bibitem[{Bautista {et~al.}(2020)Bautista, Paviot, Vargas~Magaña, de~la Torre,
  Fromenteau, Gil-Marín, Ross, Burtin, Dawson, Hou, Kneib, de~Mattia,
  Percival, Rossi, Tojeiro, Zhao, Zhao, Alam, Brownstein, Chapman, Choi,
  Chuang, Escoffier, de~la Macorra, du~Mas~des Bourboux, Mohammad, Moon,
  Müller, Nadathur, Newman, Schneider, Seo, \& Wang}]{bautista}
Bautista, J.~E., Paviot, R., Vargas~Magaña, M., {et~al.} 2020, 500, 736,
  \dodoi{10.1093/mnras/staa2800}

\bibitem[{{Colless} {et~al.}(2003){Colless}, {Peterson}, {Jackson}, {Peacock},
  {Cole}, {Norberg}, {Baldry}, {Baugh}, {Bland-Hawthorn}, {Bridges}, {Cannon},
  {Collins}, {Couch}, {Cross}, {Dalton}, {De Propris}, {Driver}, {Efstathiou},
  {Ellis}, {Frenk}, {Glazebrook}, {Lahav}, {Lewis}, {Lumsden}, {Maddox},
  {Madgwick}, {Sutherland}, \& {Taylor}}]{2dFSurvey}
{Colless}, M., {Peterson}, B.~A., {Jackson}, C., {et~al.} 2003, arXiv e-prints,
  astro, \dodoi{10.48550/arXiv.astro-ph/0306581}

\bibitem[{{Davis} {et~al.}(2021){Davis}, {Gebhardt}, {Mentuch Cooper},
  {Chisholm}, {Ciardullo}, {Farrow}, {Finkelstein}, {Gronwall}, {Gawiser},
  {Hill}, {Hopp}, {Jeong}, {Landriau}, {Liu}, {Lujan Niemeyer}, {Schneider},
  {Snigula}, \& {Tuttle}}]{Davis2021}
{Davis}, D., {Gebhardt}, K., {Mentuch Cooper}, E., {et~al.} 2021, \apj, 920,
  122, \dodoi{10.3847/1538-4357/ac1598}

\bibitem[{{Davis} {et~al.}(2023){Davis}, {Gebhardt}, {Mentuch Cooper},
  {Ciardullo}, {Fabricius}, {Farrow}, {Feldmeier}, {Finkelstein}, {Gawiser},
  {Gronwall}, {Hill}, {Hopp}, {House}, {Jeong}, {Kollatschny}, {Komatsu},
  {Landriau}, {Liu}, {Saito}, {Tuttle}, {Wold}, {Zeimann}, \&
  {Zhang}}]{Davis2022}
---. 2023, arXiv e-prints, arXiv:2301.01799.
\newblock \doarXiv{2301.01799}

\bibitem[{{Dawson} {et~al.}(2013){Dawson}, {Schlegel}, {Ahn}, {Anderson},
  {Aubourg}, {Bailey}, {Barkhouser}, {Bautista}, {Beifiori}, {Berlind},
  {Bhardwaj}, {Bizyaev}, {Blake}, {Blanton}, {Blomqvist}, {Bolton}, {Borde},
  {Bovy}, {Brandt}, {Brewington}, {Brinkmann}, {Brown}, {Brownstein}, {Bundy},
  {Busca}, {Carithers}, {Carnero}, {Carr}, {Chen}, {Comparat}, {Connolly},
  {Cope}, {Croft}, {Cuesta}, {da Costa}, {Davenport}, {Delubac}, {de Putter},
  {Dhital}, {Ealet}, {Ebelke}, {Eisenstein}, {Escoffier}, {Fan}, {Filiz Ak},
  {Finley}, {Font-Ribera}, {G{\'e}nova-Santos}, {Gunn}, {Guo}, {Haggard},
  {Hall}, {Hamilton}, {Harris}, {Harris}, {Ho}, {Hogg}, {Holder}, {Honscheid},
  {Huehnerhoff}, {Jordan}, {Jordan}, {Kauffmann}, {Kazin}, {Kirkby}, {Klaene},
  {Kneib}, {Le Goff}, {Lee}, {Long}, {Loomis}, {Lundgren}, {Lupton}, {Maia},
  {Makler}, {Malanushenko}, {Malanushenko}, {Mandelbaum}, {Manera}, {Maraston},
  {Margala}, {Masters}, {McBride}, {McDonald}, {McGreer}, {McMahon}, {Mena},
  {Miralda-Escud{\'e}}, {Montero-Dorta}, {Montesano}, {Muna}, {Myers},
  {Naugle}, {Nichol}, {Noterdaeme}, {Nuza}, {Olmstead}, {Oravetz}, {Oravetz},
  {Owen}, {Padmanabhan}, {Palanque-Delabrouille}, {Pan}, {Parejko},
  {P{\^a}ris}, {Percival}, {P{\'e}rez-Fournon}, {P{\'e}rez-R{\`a}fols},
  {Petitjean}, {Pfaffenberger}, {Pforr}, {Pieri}, {Prada}, {Price-Whelan},
  {Raddick}, {Rebolo}, {Rich}, {Richards}, {Rockosi}, {Roe}, {Ross}, {Ross},
  {Rossi}, {Rubi{\~n}o-Martin}, {Samushia}, {S{\'a}nchez}, {Sayres}, {Schmidt},
  {Schneider}, {Sc{\'o}ccola}, {Seo}, {Shelden}, {Sheldon}, {Shen}, {Shu},
  {Slosar}, {Smee}, {Snedden}, {Stauffer}, {Steele}, {Strauss}, {Streblyanska},
  {Suzuki}, {Swanson}, {Tal}, {Tanaka}, {Thomas}, {Tinker}, {Tojeiro},
  {Tremonti}, {Vargas Maga{\~n}a}, {Verde}, {Viel}, {Wake}, {Watson}, {Weaver},
  {Weinberg}, {Weiner}, {West}, {White}, {Wood-Vasey}, {Yeche}, {Zehavi},
  {Zhao}, \& {Zheng}}]{SDSSmain}
{Dawson}, K.~S., {Schlegel}, D.~J., {Ahn}, C.~P., {et~al.} 2013, \aj, 145, 10,
  \dodoi{10.1088/0004-6256/145/1/10}

\bibitem[{{de Mattia} {et~al.}(2021){de Mattia}, {Ruhlmann-Kleider},
  {Raichoor}, {Ross}, {Tamone}, {Zhao}, {Alam}, {Avila}, {Burtin}, {Bautista},
  {Beutler}, {Brinkmann}, {Brownstein}, {Chapman}, {Chuang}, {Comparat}, {du
  Mas des Bourboux}, {Dawson}, {de la Macorra}, {Gil-Mar{\'\i}n},
  {Gonzalez-Perez}, {Gorgoni}, {Hou}, {Kong}, {Lin}, {Nadathur}, {Newman},
  {Mueller}, {Percival}, {Rezaie}, {Rossi}, {Schneider}, {Tiwari}, {Vivek},
  {Wang}, \& {Zhao}}]{deMattia}
{de Mattia}, A., {Ruhlmann-Kleider}, V., {Raichoor}, A., {et~al.} 2021, \mnras,
  501, 5616, \dodoi{10.1093/mnras/staa3891}

\bibitem[{DESCollaboration {et~al.}(2021)DESCollaboration, Abbott, \&
  Aguena}]{DES}
DESCollaboration, Abbott, T. M.~C., \& Aguena, M. 2021, arXiv:2101.05765

\bibitem[{DESICollaboration(2016)}]{DESI}
DESICollaboration. 2016, arXiv:1611.00036

\bibitem[{Drake {et~al.}(2014)Drake, Graham, Djorgovski, Catelan, Mahabal,
  Torrealba, García-Álvarez, Donalek, Prieto, Williams, Larson, sen,
  Belokurov, Koposov, Beshore, Boattini, Gibbs, Hill, Kowalski, Johnson, \&
  Shelly}]{Catalina}
Drake, A.~J., Graham, M.~J., Djorgovski, S.~G., {et~al.} 2014, The
  Astrophysical Journal Supplement Series, 213, 9,
  \dodoi{10.1088/0067-0049/213/1/9}

\bibitem[{{du Mas des Bourboux} {et~al.}(2020){du Mas des Bourboux}, {Rich},
  {Font-Ribera}, {de Sainte Agathe}, {Farr}, {Etourneau}, {Le Goff}, {Cuceu},
  {Balland}, {Bautista}, {Blomqvist}, {Brinkmann}, {Brownstein}, {Chabanier},
  {Chaussidon}, {Dawson}, {Gonz{\'a}lez-Morales}, {Guy}, {Lyke}, {de la
  Macorra}, {Mueller}, {Myers}, {Nitschelm}, {Mu{\~n}oz Guti{\'e}rrez},
  {Palanque-Delabrouille}, {Parker}, {Percival}, {P{\'e}rez-R{\`a}fols},
  {Petitjean}, {Pieri}, {Ravoux}, {Rossi}, {Schneider}, {Seo}, {Slosar},
  {Stermer}, {Vivek}, {Y{\`e}che}, \& {Youles}}]{deMasDesBourboux}
{du Mas des Bourboux}, H., {Rich}, J., {Font-Ribera}, A., {et~al.} 2020, \apj,
  901, 153, \dodoi{10.3847/1538-4357/abb085}

\bibitem[{{Eisner} {et~al.}(2021){Eisner}, {Barrag{\'a}n}, {Lintott},
  {Aigrain}, {Nicholson}, {Boyajian}, {Howell}, {Johnston}, {Lakeland},
  {Miller}, {McMaster}, {Parviainen}, {Safron}, {Schwamb}, {Trouille},
  {Vaughan}, {Zicher}, {Allen}, {Allen}, {Bouslog}, {Johnson}, {Simon},
  {Wolfenbarger}, {Baeten}, {Bundy}, \& {Hoffman}}]{PlanetHunters}
{Eisner}, N.~L., {Barrag{\'a}n}, O., {Lintott}, C., {et~al.} 2021, \mnras, 501,
  4669, \dodoi{10.1093/mnras/staa3739}

\bibitem[{Farrow {et~al.}(2021{\natexlab{a}})Farrow, Sánchez, Ciardullo,
  Cooper, Davis, Fabricius, Gawiser, Gebhardt, Gebhardt, Hill, Jeong, Komatsu,
  Landriau, Chenxu, Saito, Snigula, \& Wold}]{grasshorngeb}
Farrow, D., Sánchez, A., Ciardullo, R., {et~al.} 2021{\natexlab{a}}, Monthly
  Notices of the Royal Astronomical Society, 507,
  \dodoi{10.1093/mnras/stab1986}

\bibitem[{Farrow {et~al.}(2021{\natexlab{b}})Farrow, S{\'{a} }nchez, Ciardullo,
  Cooper, Davis, Fabricius, Gawiser, Gebhardt, Gebhardt, Hill, Jeong, Komatsu,
  Landriau, Liu, Saito, Snigula, \& Wold}]{Farrow2021}
Farrow, D.~J., S{\'{a} }nchez, A.~G., Ciardullo, R., {et~al.}
  2021{\natexlab{b}}, Monthly Notices of the Royal Astronomical Society, 507,
  3187, \dodoi{10.1093/mnras/stab1986}

\bibitem[{{Gebhardt} {et~al.}(2021){Gebhardt}, {Mentuch Cooper}, {Ciardullo},
  {Acquaviva}, {Bender}, {Bowman}, {Castanheira}, {Dalton}, {Davis}, {de Jong},
  {DePoy}, {Devarakonda}, {Dongsheng}, {Drory}, {Fabricius}, {Farrow},
  {Feldmeier}, {Finkelstein}, {Froning}, {Gawiser}, {Gronwall}, {Herold},
  {Hill}, {Hopp}, {House}, {Janowiecki}, {Jarvis}, {Jeong}, {Jogee}, {Kakuma},
  {Kelz}, {Kollatschny}, {Komatsu}, {Krumpe}, {Landriau}, {Liu}, {Niemeyer},
  {MacQueen}, {Marshall}, {Mawatari}, {McLinden}, {Mukae}, {Nagaraj}, {Ono},
  {Ouchi}, {Papovich}, {Sakai}, {Saito}, {Schneider}, {Schulze},
  {Shanmugasundararaj}, {Shetrone}, {Sneden}, {Snigula}, {Steinmetz}, {Thomas},
  {Thomas}, {Tuttle}, {Urrutia}, {Wisotzki}, {Wold}, {Zeimann}, \&
  {Zhang}}]{Gebhardt21}
{Gebhardt}, K., {Mentuch Cooper}, E., {Ciardullo}, R., {et~al.} 2021, \apj,
  923, 217, \dodoi{10.3847/1538-4357/ac2e03}

\bibitem[{Gil-Marín {et~al.}(2020)Gil-Marín, Bautista, Paviot,
  Vargas-Magaña, de~la Torre, Fromenteau, Alam, Ávila, Burtin, Chuang,
  Dawson, Hou, de~Mattia, Mohammad, Müller, Nadathur, Neveux, Percival,
  Raichoor, Rezaie, Ross, Rossi, Ruhlmann-Kleider, Smith, Tamone, Tinker,
  Tojeiro, Wang, Zhao, Zhao, Brinkmann, Brownstein, Choi, Escoffier, de~la
  Macorra, Moon, Newman, Schneider, Seo, \& Vivek}]{gil-marin}
Gil-Marín, H., Bautista, J.~E., Paviot, R., {et~al.} 2020, 498, 2492,
  \dodoi{10.1093/mnras/staa2455}

\bibitem[{Land {et~al.}(2008)Land, Slosar, Lintott, Andreescu, Bamford, Murray,
  Nichol, Raddick, Schawinski, Szalay, Thomas, \&
  Vandenberg}]{SDSSotherGalaxyzoo2008}
Land, K., Slosar, A., Lintott, C., {et~al.} 2008, 388, 1686,
  \dodoi{10.1111/j.1365-2966.2008.13490.x}

\bibitem[{Laureijs {et~al.}(2011)Laureijs, Amiaux, Arduini, Auguères,
  Brinchmann, Cole, Cropper, Dabin, Duvet, Ealet, Garilli, Gondoin, Guzzo,
  Hoar, Hoekstra, Holmes, Kitching, Maciaszek, Mellier, Pasian, Percival,
  Rhodes, Criado, Sauvage, Scaramella, Valenziano, Warren, Bender, Castander,
  Cimatti, Fèvre, Kurki-Suonio, Levi, Lilje, Meylan, Nichol, Pedersen, Popa,
  Lopez, Rix, Rottgering, Zeilinger, Grupp, Hudelot, Massey, Meneghetti,
  Miller, Paltani, Paulin-Henriksson, Pires, Saxton, Schrabback, Seidel, Walsh,
  Aghanim, Amendola, Bartlett, Baccigalupi, Beaulieu, Benabed, Cuby, Elbaz,
  Fosalba, Gavazzi, Helmi, Hook, Irwin, Kneib, Kunz, Mannucci, Moscardini, Tao,
  Teyssier, Weller, Zamorani, Osorio, Boulade, Foumond, Giorgio, Guttridge,
  James, Kemp, Martignac, Spencer, Walton, Blümchen, Bonoli, Bortoletto,
  Cerna, Corcione, Fabron, Jahnke, Ligori, Madrid, Martin, Morgante, Pamplona,
  Prieto, Riva, Toledo, Trifoglio, Zerbi, Abdalla, Douspis, Grenet, Borgani,
  Bouwens, Courbin, Delouis, Dubath, Fontana, Frailis, Grazian, Koppenhöfer,
  Mansutti, Melchior, Mignoli, Mohr, Neissner, Noddle, Poncet, Scodeggio,
  Serrano, Shane, Starck, Surace, Taylor, Verdoes-Kleijn, Vuerli, Williams,
  Zacchei, Altieri, Sanz, Kohley, Oosterbroek, Astier, Bacon, Bardelli, Baugh,
  Bellagamba, Benoist, Bianchi, Biviano, Branchini, Carbone, Cardone, Clements,
  Colombi, Conselice, Cresci, Deacon, Dunlop, Fedeli, Fontanot, Franzetti,
  Giocoli, Garcia-Bellido, Gow, Heavens, Hewett, Heymans, Holland, Huang,
  Ilbert, Joachimi, Jennins, Kerins, Kiessling, Kirk, Kotak, Krause, Lahav, van
  Leeuwen, Lesgourgues, Lombardi, Magliocchetti, Maguire, Majerotto, Maoli,
  Marulli, Maurogordato, McCracken, McLure, Melchiorri, Merson, Moresco,
  Nonino, Norberg, Peacock, Pello, Penny, Pettorino, Porto, Pozzetti,
  Quercellini, Radovich, Rassat, Roche, Ronayette, Rossetti, Sartoris,
  Schneider, Semboloni, Serjeant, Simpson, Skordis, Smadja, Smartt, Spano,
  Spiro, Sullivan, Tilquin, Trotta, Verde, Wang, Williger, Zhao, Zoubian, \&
  Zucca}]{Euclid}
Laureijs, R., Amiaux, J., Arduini, S., {et~al.} 2011, Euclid Definition Study
  Report

\bibitem[{Leung {et~al.}(2017)Leung, Acquaviva, Gawiser, Ciardullo, Komatsu,
  Malz, Zeimann, Bridge, Drory, Feldmeier, Finkelstein, Gebhardt, Gronwall,
  Hagen, Hill, \& Schneider}]{Leung2017}
Leung, A.~S., Acquaviva, V., Gawiser, E., {et~al.} 2017, The Astrophysical
  Journal, 843, 130, \dodoi{10.3847/1538-4357/aa71af}

\bibitem[{Lintott {et~al.}(2008)Lintott, Schawinski, Slosar, Land, Bamford,
  Thomas, Raddick, Nichol, Szalay, Andreescu, Murray, \&
  Vandenberg}]{SDSSGalaxyZoo2008}
Lintott, C.~J., Schawinski, K., Slosar, A., {et~al.} 2008, 389, 1179,
  \dodoi{10.1111/j.1365-2966.2008.13689.x}

\bibitem[{{Mentuch Cooper} {et~al.}(2023){Mentuch Cooper}, {Gebhardt}, {Davis},
  {Farrow}, {Liu}, {Zeimann}, {Ciardullo}, {Feldmeier}, {Drory}, {Jeong},
  {Benda}, {Bowman}, {Boylan-Kolchin}, {Chavez Ortiz}, {Debski}, {Dentler},
  {Fabricius}, {Farooq}, {Finkelstein}, {Gawiser}, {Gronwall}, {Hill}, {Hopp},
  {House}, {Janowiecki}, {Khoraminezhad}, {Kollatschny}, {Komatsu}, {Landriau},
  {Lujan Niemeyer}, {Lee}, {MacQueen}, {Mawatari}, {McKay}, {Ouchi}, {Poppe},
  {Saito}, {Schneider}, {Snigula}, {Thomas}, {Tuttle}, {Urrutia}, {Weiss},
  {Wisotzki}, {Zhang}, \& {The HETDEX collaboration}}]{erincooper}
{Mentuch Cooper}, E., {Gebhardt}, K., {Davis}, D., {et~al.} 2023, arXiv
  e-prints, arXiv:2301.01826.
\newblock \doarXiv{2301.01826}

\bibitem[{Pedregosa {et~al.}(2012)Pedregosa, Varoquaux, Gramfort, Michel,
  Thirion, Grisel, Blondel, Prettenhofer, Weiss, Dubourg, VanderPlas, Passos,
  Cournapeau, Brucher, Perrot, \& Duchesnay}]{scikitlearnPedregosa11}
Pedregosa, F., Varoquaux, G., Gramfort, A., {et~al.} 2012, CoRR, abs/1201.0490

\bibitem[{Perlmutter(1999)}]{Perlmut}
Perlmutter, S. 1999, The Astronomical Journal, 517

\bibitem[{{Planck Collaboration} {et~al.}(2020){Planck Collaboration},
  {Aghanim, N.}, {Akrami, Y.}, {Ashdown, M.}, {Aumont, J.}, {Baccigalupi, C.},
  {Ballardini, M.}, {Banday, A. J.}, {Barreiro, R. B.}, {Bartolo, N.}, {Basak,
  S.}, {Battye, R.}, {Benabed, K.}, {Bernard, J.-P.}, {Bersanelli, M.},
  {Bielewicz, P.}, {Bock, J. J.}, {Bond, J. R.}, {Borrill, J.}, {Bouchet, F.
  R.}, {Boulanger, F.}, {Bucher, M.}, {Burigana, C.}, {Butler, R. C.},
  {Calabrese, E.}, {Cardoso, J.-F.}, {Carron, J.}, {Challinor, A.}, {Chiang, H.
  C.}, {Chluba, J.}, {Colombo, L. P. L.}, {Combet, C.}, {Contreras, D.},
  {Crill, B. P.}, {Cuttaia, F.}, {de Bernardis, P.}, {de Zotti, G.},
  {Delabrouille, J.}, {Delouis, J.-M.}, {Di Valentino, E.}, {Diego, J. M.},
  {Dor\'e, O.}, {Douspis, M.}, {Ducout, A.}, {Dupac, X.}, {Dusini, S.},
  {Efstathiou, G.}, {Elsner, F.}, {En\ss{}lin, T. A.}, {Eriksen, H. K.},
  {Fantaye, Y.}, {Farhang, M.}, {Fergusson, J.}, {Fernandez-Cobos, R.},
  {Finelli, F.}, {Forastieri, F.}, {Frailis, M.}, {Fraisse, A. A.},
  {Franceschi, E.}, {Frolov, A.}, {Galeotta, S.}, {Galli, S.}, {Ganga, K.},
  {G\'enova-Santos, R. T.}, {Gerbino, M.}, {Ghosh, T.}, {Gonz\'alez-Nuevo, J.},
  {G\'orski, K. M.}, {Gratton, S.}, {Gruppuso, A.}, {Gudmundsson, J. E.},
  {Hamann, J.}, {Handley, W.}, {Hansen, F. K.}, {Herranz, D.}, {Hildebrandt, S.
  R.}, {Hivon, E.}, {Huang, Z.}, {Jaffe, A. H.}, {Jones, W. C.}, {Karakci, A.},
  {Keih\"anen, E.}, {Keskitalo, R.}, {Kiiveri, K.}, {Kim, J.}, {Kisner, T. S.},
  {Knox, L.}, {Krachmalnicoff, N.}, {Kunz, M.}, {Kurki-Suonio, H.}, {Lagache,
  G.}, {Lamarre, J.-M.}, {Lasenby, A.}, {Lattanzi, M.}, {Lawrence, C. R.}, {Le
  Jeune, M.}, {Lemos, P.}, {Lesgourgues, J.}, {Levrier, F.}, {Lewis, A.},
  {Liguori, M.}, {Lilje, P. B.}, {Lilley, M.}, {Lindholm, V.},
  {L\'opez-Caniego, M.}, {Lubin, P. M.}, {Ma, Y.-Z.}, {Mac\'{\i}as-P\'erez, J.
  F.}, {Maggio, G.}, {Maino, D.}, {Mandolesi, N.}, {Mangilli, A.},
  {Marcos-Caballero, A.}, {Maris, M.}, {Martin, P. G.}, {Martinelli, M.},
  {Mart\'{\i}nez-Gonz\'alez, E.}, {Matarrese, S.}, {Mauri, N.}, {McEwen, J.
  D.}, {Meinhold, P. R.}, {Melchiorri, A.}, {Mennella, A.}, {Migliaccio, M.},
  {Millea, M.}, {Mitra, S.}, {Miville-Desch\^enes, M.-A.}, {Molinari, D.},
  {Montier, L.}, {Morgante, G.}, {Moss, A.}, {Natoli, P.},
  {N\o{}rgaard-Nielsen, H. U.}, {Pagano, L.}, {Paoletti, D.}, {Partridge, B.},
  {Patanchon, G.}, {Peiris, H. V.}, {Perrotta, F.}, {Pettorino, V.},
  {Piacentini, F.}, {Polastri, L.}, {Polenta, G.}, {Puget, J.-L.}, {Rachen, J.
  P.}, {Reinecke, M.}, {Remazeilles, M.}, {Renzi, A.}, {Rocha, G.}, {Rosset,
  C.}, {Roudier, G.}, {Rubi\~no-Mart\'{\i}n, J. A.}, {Ruiz-Granados, B.},
  {Salvati, L.}, {Sandri, M.}, {Savelainen, M.}, {Scott, D.}, {Shellard, E. P.
  S.}, {Sirignano, C.}, {Sirri, G.}, {Spencer, L. D.}, {Sunyaev, R.},
  {Suur-Uski, A.-S.}, {Tauber, J. A.}, {Tavagnacco, D.}, {Tenti, M.},
  {Toffolatti, L.}, {Tomasi, M.}, {Trombetti, T.}, {Valenziano, L.},
  {Valiviita, J.}, {Van Tent, B.}, {Vibert, L.}, {Vielva, P.}, {Villa, F.},
  {Vittorio, N.}, {Wandelt, B. D.}, {Wehus, I. K.}, {White, M.}, {White, S. D.
  M.}, {Zacchei, A.}, \& {Zonca, A.}}]{plank}
{Planck Collaboration}, {Aghanim, N.}, {Akrami, Y.}, {et~al.} 2020, A\&A, 641,
  A6, \dodoi{10.1051/0004-6361/201833910}

\bibitem[{Raddick {et~al.}(2019)Raddick, Prather, \& Wallace}]{pratherWallace}
Raddick, M.~J., Prather, E.~E., \& Wallace, C.~S. 2019, 28, 636,
  \dodoi{10.1177/0963662519840222}

\bibitem[{Riess {et~al.}(2021)Riess, Casertan, Yuan, \& Bowers}]{Riess21}
Riess, A.~G., Casertan, S., Yuan, W., \& Bowers, J.~B. 2021, The Astronomical
  Journal Letters, 908

\bibitem[{Riess {et~al.}(1998)Riess, Filippenko, V., Challis, \&
  Clocchiatti}]{Riess98}
Riess, A.~G., Filippenko, V., A., Challis, P., \& Clocchiatti, A. 1998, The
  Astronomical Journal, 116, 1009

\bibitem[{Santos-Fernandez \& Mengersen(2021)}]{Santos2021}
Santos-Fernandez, E., \& Mengersen, K. 2021, Methods in Ecology and Evolution,
  12, 1533, \dodoi{https://doi.org/10.1111/2041-210X.13623}

\bibitem[{Sánchez {et~al.}(2014)Sánchez, Montesano, Kazin, Aubourg, Beutler,
  Brinkmann, Brownstein, Cuesta, Dawson, Eisenstein, Ho, Honscheid, Manera,
  Maraston, {McBride}, Percival, Ross, Samushia, Schlegel, Schneider, Skibba,
  Thomas, Tinker, Tojeiro, Wake, Weaver, White, \& Zehavi}]{Sanchez14}
Sánchez, A.~G., Montesano, F., Kazin, E.~A., {et~al.} 2014, 440, 2692,
  \dodoi{10.1093/mnras/stu342}

\bibitem[{Tegmark {et~al.}(2004)Tegmark, Strauss, Blanton, Abazajian, Dodelson,
  Sandvik, Wang, Weinberg, Zehavi, Bahcall, Hoyle, Schlegel, Scoccimarro,
  Vogeley, Berlind, Budavari, Connolly, Eisenstein, Finkbeiner, Frieman, Gunn,
  Hui, Jain, Johnston, Kent, Lin, Nakajima, Nichol, Ostriker, Pope, Scranton,
  Seljak, Sheth, Stebbins, Szalay, Szapudi, Xu, Annis, Brinkmann, Burles,
  Castander, Csabai, Loveday, Doi, Fukugita, Gillespie, Hennessy, Hogg,
  Ivezi\ifmmode~\acute{c}\else \'{c}\fi{}, Knapp, Lamb, Lee, Lupton, McKay,
  Kunszt, Munn, O'Connell, Peoples, Pier, Richmond, Rockosi, Schneider,
  Stoughton, Tucker, Vanden~Berk, Yanny, \& York}]{WMAP}
Tegmark, M., Strauss, M.~A., Blanton, M.~R., {et~al.} 2004, Phys. Rev. D, 69,
  103501, \dodoi{10.1103/PhysRevD.69.103501}

\bibitem[{van~der Maaten(2015)}]{Maaten2015}
van~der Maaten, L. 2015, Journal of Machine Learning Research, 15, 3221

\bibitem[{van~der Maaten \& Hinton(2008)}]{vandermaaten08}
van~der Maaten, L., \& Hinton, G. 2008, Journal of Machine Learning Research,
  9, 2579

\bibitem[{Weinberg(1989)}]{Weinberg}
Weinberg, S. 1989, Rev. Mod. Phys., 61, 1, \dodoi{10.1103/RevModPhys.61.1}

\end{thebibliography}

\end{document}